\documentclass[letter,oldversion]{aa}
\usepackage{graphicx}
\usepackage{txfonts}

\newcommand{\micron} {$\mu$m}
\newcommand\arcdeg{\mbox{$^\circ$}}
\newcommand{\kms}{km~s$^{-1}$}

\newcommand{\ms}{\mbox{$M_{\odot}$}}

\newcommand{\cgsint}{erg\,s$^{-1}$\,cm$^{-2}$\,sr$^{-1}$}
\newcommand{\cgsflx}{erg\,s$^{-1}$\,cm$^{-2}$}
\makeatletter
 \renewcommand\section{\@startsection {section}{1}{\z@}%
     {-2.5ex \@plus -1ex \@minus -.2ex}%
     {1.3ex \@plus.2ex}%
    {\centering\bfseries}}
\makeatother

\begin{document}
   \title{Probing the Mass and Structure of the Ring Nebula in Lyra with SOFIA/GREAT Observations of the [CII]\,158\,micron~Line}

   \subtitle{}

   \author{R. Sahai\inst{1}
          \and
         M. R. Morris\inst{2}        
          \and
         M. W. Werner\inst{1} 
          \and
         R. G{\"u}sten\inst{3}          
           \and
         H. Wiesemeyer\inst{3}         
           \and
         G. Sandell\inst{4}
          }
   \institute{Jet Propulsion Laboratory,   
              4800 Oak Grove Drive, Pasadena, CA 91109-8001, USA, 
              \email{raghvendra.sahai@jpl.nasa.gov}
         \and
             University of California, Los Angeles, CA 90095-1547, USA
         \and
             Max-Planck-Institut f{\"u}r Radioastronomie, Auf dem H{\"u}gel 69, 53121 Bonn, Germany
         \and
             SOFIA-USRA, NASA Ames Research Center, MS 232-12, Building N232/Rm. 146, Moffett Field, CA 94035-0001, USA
            }

   \date{Received ; accepted }

 
  \abstract
  {We have obtained new velocity-resolved spectra of the [CII]\,158\,\micron~line towards the Ring Nebula in Lyra (NGC\,6720), one of the best-studied planetary nebulae, in order to probe its controversial 3-dimensional structure and to estimate the mass of circumstellar material in this object.
We used the Terahertz receiver GREAT aboard the SOFIA airborne telescope to obtain the [CII] spectra at eight locations within and 
outside the bright optical ring of NGC\,6720. Emission was detected at all positions except for the most distant position along the nebula's minor axis, and 
generally covers a broad velocity
range, $\Delta$V$\sim50$\,\kms~(FWZI), except at a position along the major axis located just outside the optical ring, where it is significantly 
narrower ($\Delta$V$\sim25$\,\kms).  
The one narrow spectrum appears to be probing circumstellar material lying outside the main
nebular shell that has not been accelerated by past fast wind episodes from the central star, and therefore most likely comes from
equatorial and/or low-latitude regions of this bipolar nebula. 
Along lines-of-sight passing within about 10\arcsec~of the nebular center, the CII column density is a factor 46  
higher than the CO column density. 
The total mass of gas associated with the [CII] emission inside a circular region of 
diameter $87\farcs 5$~is at least 0.11\ms. 
A significant amount of [CII] flux arises from a photodissociation region immediately outside the bright optical ring, 
where we find a CII to CO ratio of $>6.5$, lower than that seen 
towards the central region. Comparing our data with lower-quality CI spectra, which indicate similarly large CI/CO ratios in NGC\,6720, 
we conclude that the bulk of elemental carbon in NGC\,6720 is divided roughly equally between CII and CI, and that the emissions  
from these species are far more robust tracers 
of circumstellar material than CO in this object and other evolved planetary nebulae.
}

\titlerunning{[CII]\,158\,micron~in NGC 6720}
   \maketitle

%

\section{Introduction}
\vskip -0.13in
Planetary Nebulae (PNe) result when ``ordinary" stars, with main-sequence masses 1-8\ms, die extraordinary deaths. During their
post-main-sequence evolutionary phase (i.e.,
as AGB stars), these stars eject half or more of their total mass in the form of
nucleosynthetically-enriched material into the interstellar medium (ISM) -- a process which
dramatically alters the course of stellar evolution, and plays a key role in the chemical evolution of the Galaxy. PNe
are formed when most of the stellar envelope has been ejected and the hot central star's radiation field ionizes the
ejecta. 

PNe are almost a unique astrophysical class in terms of being bright at all wavelength bands from X-ray to radio,
and are ideal laboratories for investigating important
astrophysical phenomena such as nucleosynthesis, (magneto)hydrodynamics of colliding winds, astrochemistry,
photodissociation, and the evolution of dust grains (via UV and shock-processing) because their geometries and 
kinematics are generally believed to be well understood (e.g., Gurzadyan 1997). The history of the mass-loss processes which result in a PN,
beginning from its life as an AGB star, is written in the spatio-kinematic structure of the circumstellar ejecta surrounding
the central star. Different geometries can require radically different formation models -- for example, in the binary-driven
model for aspherical PNe, ellipsoidal shapes result from sub-stellar mass companions, whereas bipolar
shapes require interaction with a stellar-mass companion (Soker 1996).

However, inferring the 3D physical structure of a PN from its observed 2D projection on the sky can be a difficult 
task. This is because PNe appear in a dazzling variety of complex morphologies, with collimated, bipolar and multipolar
shapes being dominant, and round ones being rather rare, as clearly seen in imaging surveys with HST  
(e.g., Sahai \& Trauger 1998; Sahai, Morris \& Villar 2011). Furthermore, although PNe shapes have traditionally 
been defined by optical imaging, the latter typically do not show where
the bulk of the mass in these objects lies: 
in a photodissociation region (PDR) surrounding the bright optical nebula, of up to $\sim$1\,\ms, compared to
$\sim$0.1\,\ms~in ionized gas (c.f., Hollenbach \& Tielens 1999). This PDR may represent most of the mass
ejected during the AGB phase; some fraction of it is sculpted and compressed by fast post-AGB winds from the central star, resulting in
the complex shapes of PNe seen in optical imagery. The bulk of the PDR material radiates most of its energy via thermal emission in
the far-infrared.

At a distance of 0.7\,kpc, NGC\,6720 (M\,57), the Ring Nebula in Lyra, is among the brightest and most extensively studied
PNe (e.g., O'Dell et al. 2007:\,ODell07, and references therein, van Hoof et al. 2010). NGC6720 has an elliptical ring shape
(size $\sim90\arcsec\times60\arcsec$) in optical emission lines. This bright ring lies at the center of a 
large ``double" halo structure: a brighter inner one (size $\sim160\arcsec\times146\arcsec$) consisting of a system of limb-brightened loops, and an outer 
faint one which is irregular, knotty and round ($\sim230$\arcsec~diameter: Balick et al. 1992). Although it is considered a ``classical" 
nebula because of the bright ring's simple 2D shape, its detailed 3D structure remains controversial,
in spite of many studies undertaken to elucidate it (see, e.g., ODell07). 

We report here the results of a small program conducted as part of Basic Science observations on the Stratospheric
Observatory For Infrared Astronomy (SOFIA), 
to use velocity-resolved observations of [CII]\,158\,\micron~line emission for probing the
mass and 3-D structure of NGC\,6720. The use of the [CII]\,158\,micron~line as a probe stemmed from our expectation that, because [CII] 
is present both in the ionized and atomic components of the circumstellar environment, and 
because this line has a low critical density ($\lesssim\,\mathrm{few}\,10^3\,$cm$^{-3}$), it should be readily observable
both within and outside the optically-bright PN ring.

\vspace{-0.05in}
\section{Observations}
\begin{figure*}
\centering
\includegraphics[width=15.0cm,clip]{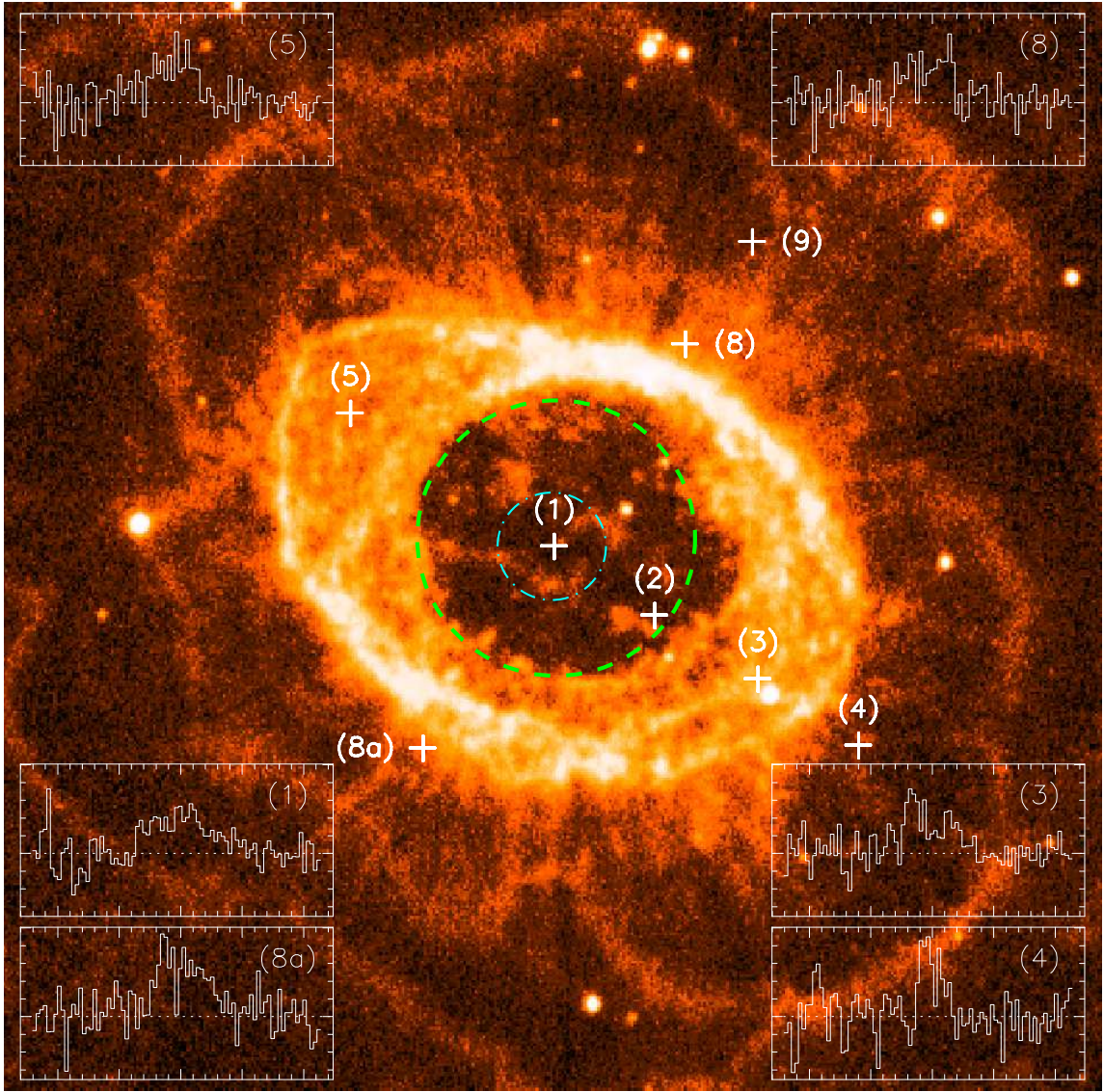}
\caption{Spectra of the [CII]158\,\micron~emission observed with GREAT towards NGC\,6720 obtained at different 
offsets from the center (see Table\,\ref{tabflx}), overlaid on a 
$157\farcs 9 \times 157\farcs 9$ H$_2$ 2.1\,\micron~image (taken by David Thompson). The 
positions of the $15\farcs 6$~GREAT beam, shown in cyan at the center (J2000 RA=18:53:35.08, DEC=33:01:45.0) are marked by crosses. 
The small tick marks are separated by 10\,\kms~on the velocity axis ($V_{lsr}$), and by 0.1\,K on the intensity axis ($T_A^*$). 
The spectrum at position \#4 has been scaled by a factor 0.75. The equator of the multipolar model we propose for NGC\,6720 is shown by the 
green dashed circle with diameter 40\arcsec.
}
\vskip -0.15in
\label{n6720-img-spec}
\end{figure*}

\vskip -0.05in
The observations were performed with the (dual-channel) German Receiver for
Astronomy at Terahertz Frequencies (GREAT\footnote{GREAT is a development by
the MPI f{\"u}r Radioastronomie and the Universit{\"a}t zu K{\"o}ln, in cooperation with the MPI f{\"u}r Sonnensystemforschung 
and the DLR Institut f{\"u}r Planetenforschung.}: Heyminck et al.
2012) on board SOFIA during three flights with durations of  70, 85 and 60 minutes on 2011 July 12, July 17, and Sept 28, at altitudes of 43000, 43000 and 41000 ft, respectively. We used the
Fast Fourier Transform spectrometer with 1.5\,GHz instantaneous bandwidth (236.6\,\kms~at the [CII] line frequency of 1900.5369\,GHz). 
The symmetric beam-switching mode was employed with a throw of 6\arcmin~along the Dec axis. 
Pointing was accurate to better than 5\arcsec, and the 
system temperature was $\sim$3950--4530\,K. 
Spectra of the [CII] line were obtained at eight locations inside 
and outside NGC\,6720's bright ring along its major and 
minor axes (Fig.\,\ref{n6720-img-spec}), with an angular resolution of $\sim15\farcs 6$. 
The data were processed with the latest version of the GREAT calibration pipeline.
We reduced the spectra further by boxcar smoothing these to a resolution of 2.888\,\kms~and removal of linear baselines. 
The typical integration time (1$\sigma$ noise) per position (i.e., excluding the time spent on the ``off" positions) 
ranged between 5--8\,min (0.06--0.1\,K), except for \#4, where it was 3.5\,min (0.14\,K). 
Using the beam efficiencies $\eta_c=0.51$ for [CII] and the forward efficiency $\eta_f=0.95$ 
(Heyminck et al. 2012), all data were converted to a main-beam brightness temperature scale 
$T_{mb} = \eta_f\,T_A^*/\eta_c$ for computing the total [CII] flux.  The 
temperature scale is uncertain by 20\%. 

\vspace{-0.05in}
\section{Results}
\vskip -0.05in
We detected emission at all positions except \#9, where the rms noise is 0.09\,K (Table\,\ref{tabflx}). The emission
generally covers a broad velocity
range, $\Delta$V$\sim$50\,\kms (full-width at zero intensity: FWZI), except at position \#4, where it is significantly narrower 
($\Delta$V$\sim25-30$\,\kms) (Fig.\,2). An average spectrum from the central (\#1) and four symmetric locations on the major and 
minor axes (\#3,\#5,\#8,\#8a) shows a symmetric profile with a central double-peaked core of width $\Delta$V$\sim48$\,\kms, 
centered at $V_{lsr}=-2.6$\,\kms, and a weaker high-velocity pedestal with a total width of about 100\,\kms~at its base. The observed 
width of the high-velocity component is limited by the signal-to-noise, and is likely larger than 100\,\kms~(e.g., by folding and averaging 
this profile around its center velocity, we reduce the noise and find clear evidence for emission extending over 120\,\kms). The line profile 
towards position \#1, although noisy, clearly does not show even a hint of a minimum around the systemic velocity. 
Since an expanding hollow shell would produce a strong minimum, we conclude that C$^+$ is not absent from the region interior to the bright ring.  

We compute the [CII] column density at different 
locations in the nebula from the observed [CII] intensity, $I([CII])$ (\cgsint), using 
equation A.5 of Schneider et al. (2003), $N(CII)=6.3\times10^{20}\,I([CII])$\,cm$^{-2}$ (Table\,\ref{tabflx}). We have 
reasonably assumed 
that the density and temperature are high enough for thermalized
emission -- the upper level of the [CII] line is 91\,K above ground, and 
the critical densities for collisional excitation by electrons (at $T_e=8000$\,K), H, 
and H$_2$ are 46, $3\times10^3$ and $5\times10^3$\, cm$^{-3}$, respectively (Stacey et al. 2010, Schneider et al. 2003), 
to be compared with the electron density 
in the ionized nebula (500--700\,cm$^{-3}$, ODell07), and the density ($10^5$\,cm$^{-3}$) and temperature (300\,K) 
in the PDR of NGC\,6720 (Liu et al. 2001). 
Hence, with $I([CII])=2\times10^{-4}$\,\cgsint~towards the nebular center, we find 
$N(CII)=1.2\times10^{17}$\,cm$^{-2}$.

\begin{table}[]
\caption{GREAT Observations of [CII] in NGC\,6720}
\begin{center}
\begin{tabular}{llll}
\hline
Position & Offset & F([CII]) & N(CII) \\
Label & $\Delta\alpha$(\arcsec),\,$\Delta\delta$(\arcsec) & K\,.\,km\,s$^{-1}$ & $10^{17}$\,cm$^{-2}$\\
\hline
1 & (0,\,0)$^{\mathrm{a}}$ & 27.9  & 1.24  \\
2 & ($-14.7,\,-9.5$)      & 15.0  & 0.67  \\
3 & ($-29.3,\,-19.1$)     & 21.8  & 0.97  \\
5 & ($+29.3,\,+19.1$)     & 29.2  & 1.29  \\
8 & ($-19.1,\,+29.3$)     & 22.2  & 0.98  \\
8a& ($+19.1,\,-29.3$)     & 41.2  & 1.82  \\
4 & ($-44.0,\,-28.6$)     & 12.5  & 0.55  \\
9 & ($-28.6,\,+44.0$)     & $<4.5$ & 0.20  \\
\hline
\end{tabular}
\begin{list}{}{}
\item[$^{\mathrm{a}}$]coordinates are J2000 RA=18:53:35.08, DEC=33:01:45.0
\end{list}
\vspace{-3.5em}
\end{center}
\label{tabflx}
\end{table}

Since we have sampled the nebula sparsely with beams separated by $17\farcs 5$, 
we estimate the total [CII] flux by approximating  
the [CII] emitting region as a central, circular region of $17\farcs 5$~diameter, surrounded by two concentric annuli. 
The mean radius of annulus 1 (2) is equal to the offsets of position \#2 (\#3; same for \#'s 5,8 \& 8a) from the center, 
and each has a width of $17\farcs 5$. We derive the average intensity in 
each region from the average of the spectra sampling it, and multiply it by the region's area to derive the flux. 
We thus find that the total [CII] flux, $F([CII])=2.54\times10^{-11}$\,\cgsflx. Assuming all of the carbon is in C$^{+}$, that the 
distance is 0.7\,kpc, and taking 
$C/H=7.9\times10^{-4}$ for NGC\,6720 (Liu et al. 2001, Table 5), we find 
that the total mass of gas associated with the [CII] emission inside a circular region of 
diameter $87\farcs 5$~is at least 0.114\ms. 
Our mass estimate is a very conservative lower limit because (i) we have not taken into account the emission from locations 
beyond the main ring, such as position \#4, and (ii) a comparable fraction of carbon is likely to be neutral (see below), 
We note that Liu et al. (2001)
estimated a mass of 0.03\ms~from their [CII]\,158\,\micron~line flux of $6.8\times10^{-12}$\,\cgsflx~for NGC\,6720, 
measured using the ISO Long Wavelength Spectrometer (LWS) with
a single $\sim$70\arcsec~aperture covering the main optical nebula. 
Clearly, a significant amount of [CII] flux lies outside the 70\arcsec~ISO/LWS aperture, i.e., 
in the PDR beyond the main optical nebula.

We compare our [CII] results with those from studies of CO and CI emission in 
NGC\,6720. For CO, Bachiller at al. (1989: BBMG89) have obtained complete maps in the J=1--0 and 2--1 lines, 
whereas for CI, Bachiller at al. (1994: BHCF94) 
detect emission in the 609\,\micron~line with low S/N at a few (3 of 5) positions along the minor axis. 
The CO and CI column densities towards the nebular
center are $N(CO)=2.6\times10^{15}$\,cm$^{-2}$ (BBMG89) and $N(CI)\sim10^{17}$\,cm$^{-2}$ (BHCF94). Hence, both the CII/CO and the CI/CO 
abundance ratios are about 50 
towards the nebular center, i.e., almost all of the nebular carbon is distributed roughly equally between C$^{+}$ and C. 
Towards positions offset from the nebular center, the CII/CO abundance ratio, while still large, is not as extreme.    
We find (i) $N(CII)=5.5\times10^{16}$\,cm$^{-2}$ towards position \#4, and (ii)  
$N(CO)=8.4\times10^{15}$\,cm$^{-2}$ near position \#4 from BBMG89's CO J=2--1 map, using their spectrum at offset position 
($\Delta\alpha,\Delta\delta=-40\arcsec,-20\arcsec$). Since this spectrum (obtained with a 13\arcsec~beam) samples the strongest  
emission peak in the CO map, whereas our \#4 spectrum ($15\farcs 6$~beam), is offset by $9\farcs 5$~from that peak,  
our derived C$^{+}$ to CO ratio here is at least 6.5. 
Our CII study strongly supports the hypothesis that the CO emission region in NGC\,6720 is really 
a PDR (BHCF94). We note that the CO J=2--1 emission shows significant clumpy structure, and given the 
strong [CII] emission seen at position \#4 where the CO emission is also strong, and the lack of [CII] 
emission at \#9 (3$\sigma$ upper limit of $N([CII])<2\times10^{16}$\,cm$^{-2}$ assuming same velocity width as at position \#4), 
where the CO emission is very weak, it is likely that the [CII] is similarly clumpy. 

Noting that the CII column densities at all positions we have 
sampled in NGC\,6720 (Table\,\ref{tabflx}) exceed the maximum CO column density in NGC\,6720 by large factors, 
and that the average CI/CO ratio is 10 (BHCF94), we conclude that CII and CI are far more robust tracers of 
circumstellar mass in this object than CO. If we assume that the abundances of CII and CI are roughly equal, then 
our circumstellar mass estimate from [CII] data increases to $>0.23$\ms. Hence, given estimates of the current central star mass 
of 0.61--0.62\,\ms~(Odell07), and that the 
mass of the outer halo has not been accounted for, it is likely that the mass of NGC\,6720's progenitor star exceeded 1\,\ms.

\vspace{-0.1in}
\subsection{Structure}
\vskip -0.1in
Models for the 3-D structure of NGC\,6720 can be divided into two broad classes.
The first (model 1) is a (roughly) prolate ellipsoid, having open ends along the long axis: thus a barrel shape, with an equatorial 
density enhancement and the axis of symmetry inclined at 30\arcdeg~to the line of sight (Masson 1990, Guerrero et al. 1997: GMC97).  
This model has been refined by Hiriart (2004) and ODell07 (using mapping of the 2.1\,\micron~H$_2$ line, 
and multislit optical spectra, respectively) to one in which the barrel cross-section is elliptical and it is 
seen nearly pole-on. 
In the second class of models (model 2), NGC\,6720 is seen
as a nearly pole-on bipolar nebula based on multislit spectra along the major axis (Bryce et al. 1994).
Kwok et al. (2008) compare deep emission-line images of NGC\,6720 with another
well-studied PN, NGC\,6853, and suggest that both have the same biconical 3-D structure, except that NGC\,6853 is oriented
edge-on, whereas NGC\,6720 is nearly pole-on. In the Kwok et al. (2008) model, the inner and outer halo together represent the 
projections of three pairs of bi-conical outflows with similar opening angles, viewed along their common symmetry axes. From their CO J=2-1 study, 
BBMG89 infer a structure which has elements of both class 1 and class 2: a hollow cylindrical shape with its axis inclined to the line-of-sight. 

We propose that NGC\,6720 is a PN with   
a barrel-shaped central region viewed (nearly) along its axis, which can reconcile the above model classes. 
Barrel-shaped central regions have recently been recognized as an 
important feature of many PNe in a comprehensive new morphological classification system based on HST images of 
119 young PNe by Sahai, Morris \& Villar (2011). NGC\,6720 is also multipolar, with the ``flower-petal" structures in the H$_2$ and [NII] images 
seen in the ``inner halo" of NGC\,6720 being the limb-brightened peripheries of these lobes.  The presence of such lobes  
is also indicated by the appearance of multiple, faint, red- and blue-shifted velocity components within the overall position-velocity ellipse seen in optical long-slit spectra reported by GMC97 and interpreted as 
resulting from ``bubbles of material (that) protrude from the main nebula". 
The NGC\,6720 ring shows several extended filamentary structures which suggests that a significant fraction of the ring's optical line emission comes from the bright basal regions of multiple lobes emanating from the central region; the continuation of these regions to slightly higher latitudes is seen in 
the fainter, more diffuse, emission in the H$_2$ image immediately outside the bright ring (Fig.\,\ref{n6720-img-spec}). This suggestion finds support in GMC97's finding that ``the velocity structure of the edge of the 
nebula shows a complex behaviour",  ``reflecting the projected velocity of these clumps and bubbles". Optical spectra 
of the flower-petal region (Fig. 8 of GMC97) show distinct narrow
components separated by $\sim25-30$\,\kms. In our model, these result from the inclined walls of
lobes on the near- and far-side of the equatorial plane.
These spectra also show the presence
of point-symmetric pairs of knotty structures which are
consistent with an origin in oppositely-directed multiple outflows,
rather than bubbles created by a fast wind flowing through a
fragmented shell as GMC97 propose, because it is very unlikely
that a stochastic process (fragmentation) can create
point-symmetrically located holes in a shell.
The multipolar PN He\,2-47 (Sahai 2000) provides us a rough idea of what NGC\,6720's lobes might look like if 
the symmetry axis of its central barrel was in (or nearly in) the sky-plane.

Although a detailed interpretation of all the published data on NGC\,6720 with our proposed model is outside the scope of this paper, 
our [CII] data appear consistent with this model, provided the lobes are expanding at similar velocities. 
In this model, the [CII] emission at positions 
3, 5, 8, and 8A, which likely includes emission from both the barrel-shaped central region and low-latitude regions of 
multipolar lobes, is expected to be roughly similar, as observed.
The spectrum at position \#4 provides insight into the circumstellar material that lies outside the main
nebular ring and whose physical properties are not well known -- the narrow profile ($\Delta$V$\sim25$\,\kms) 
suggests that the [CII] emission is probing
material that has not been accelerated by past fast wind episodes from the central star, and therefore most likely comes from
equatorial and low-latitude regions of the slowly-expanding progenitor AGB wind. In contrast, the emission from positions \#8 and \#8a which, like that from \#4, also samples material largely just
outside the shell's periphery but along the minor axis, is broad ($\Delta$V$\sim40-45$\,kms) and may be associated with
the multipolar lobes whose peripheries show up in projection as the ``flower-petals" in the H$_2$ image (Fig.\,\ref{n6720-img-spec}). 

In summary, our results show that for NGC\,6720 (and by inference other evolved PNe), C$^{+}$  
is a much more robust tracer of the circumstellar material than CO because it is far more abundant. We therefore 
expect future GREAT [CII] observations to play a major role in the study of evolved PNe. 

\begin{figure}
\resizebox{0.42\textwidth}{!}{\includegraphics{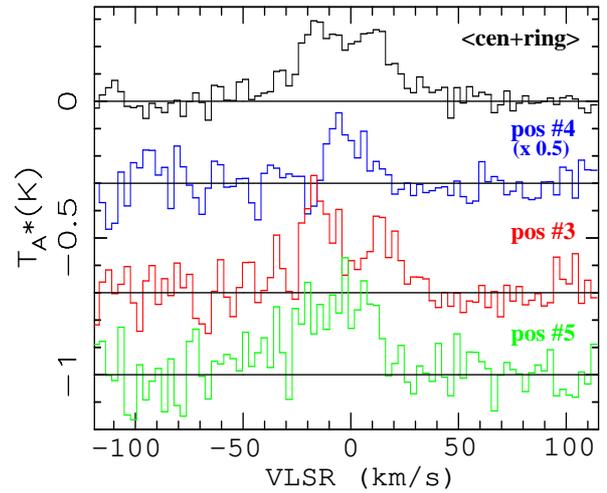}}
{\caption{[CII] spectra obtained at specific locations in NGC\,6720 with GREAT. The \#4 (rescaled by 0.5),  \#3, and \#5 spectra have been shifted vertically by -0.3, -0.7, and -1.0\,K or clarity. The ``cen+ring" spectrum is an average of broad spectra
found at the central (position \#1) and four symmetric locations on the major and minor axis (positions \#3,\#5,\#8,\#8a).}
}
\vskip -0.15in
\label{n6720spec}
\end{figure}

\begin{acknowledgements}
\vskip -0.1in     
This study is based on observations made with the NASA/DLR Stratospheric Observatory for Infrared Astronomy (SOFIA). 
SOFIA Science Mission Operations is operated by the Universities Space Research Association, Inc., under NASA contract NAS2-97001, 
and the Deutsches SOFIA Institut under DLR contract 50\,OK\,0901. RS's contribution to the
research described here was carried out at the Jet Propulsion Laboratory, California Institute of Technology, under a
contract with NASA. Financial support was provided by NASA through a Long Term Space Astrophysics
award to RS and MM, and a SOFIA award to RS.

\end{acknowledgements}

\end{document}